\documentclass{article}
\usepackage[T1]{fontenc}
\usepackage{wrapfig}
\usepackage[portuges,english]{babel}
\usepackage{amssymb,latexsym,amsmath,color,mathrsfs,ifsym,graphics,stmaryrd} 
\usepackage[colorlinks,linkcolor=blue,urlcolor=blue,citecolor=black,
plainpages=false,pdfpagelabels,breaklinks]{hyperref}

\title{Hilbert Space Quantum Mechanics is Contextual\\(Reply to R. B. Griffiths)}

\author{{\sc Christian de Ronde}\thanks{Fellow Researcher of the Consejo
Nacional de Investigaciones Cient\'{\i}ficas y T\'ecnicas. E-mail: cderonde@vub.ac.be}}
\date{}

\begin{document}
\maketitle

\begin{center}
\begin{small}
CONICET, Buenos Aires University - Argentina \\
Center Leo Apostel and Foundations of  the Exact Sciences\\
Brussels Free University - Belgium \\
\end{small}
\end{center}

\begin{abstract}
\noindent In a recent paper Robert Griffiths \cite{Griffiths13} has argued, based on the consistent histories interpretation, that Hilbert space quantum mechanics (QM) is non-contextual. According to Griffiths, the problem of contextuality disappears if the apparatus is ``designed and operated by a competent experimentalist'' and we accept the {\it Single Framework Rule (SFR)}. In the present paper we will argue that Griffiths' conclusion is untenable. Firstly, we will argue that Kochen-Specker (KS) type contextuality has nothing to do with the measurement processes nor the measurement problem ---solving this problem does not necessarily remove contextuality. Secondly, that the consideration of counterfactual situations is endemic when reasoning by means of physics. Griffiths' {\it SFR} amounts to an unprincipled (or impractical) denial of this fact. Thirdly, that Griffiths' demonstration implies, as a direct consequence, that QM allows a {\it Global Valuation} for all properties independently of the context. This would mean, not only that KS theorem is false, but also that the {\it SFR} is completely unnecessary. Finally, we will argue that quantum contextuality should be reconsidered ---instead as a problem that we need to escape--- as a main feature of the formalism which must be positively taken into account in order to provide a coherent representation of physical reality and explain what is QM talking about. 
\end{abstract}

\renewenvironment{enumerate}{\begin{list}{}{\rm \labelwidth 0mm
\leftmargin 0mm}} {\end{list}}

\newcommand{\ita}{\textit}
\newcommand{\mcal}{\mathcal}
\newcommand{\mfrak}{\mathfrak}
\newcommand{\mbb}{\mathbb}
\newcommand{\mrm}{\mathrm}
\newcommand{\msf}{\mathsf}
\newcommand{\mscr}{\mathscr}
\newcommand{\lra}{\leftrightarrow}
\renewenvironment{enumerate}{\begin{list}{}{\rm \labelwidth 0mm
\leftmargin 5mm}} {\end{list}}

\newtheorem{dfn}{\sc{Definition}}[section]
\newtheorem{thm}{\sc{Theorem}}[section]
\newtheorem{lem}{\sc{Lemma}}[section]
\newtheorem{cor}[thm]{\sc{Corollary}}
\newcommand{\Proof}{\textit{Proof:} \,}
\newcommand{\cqd}{{\rule{.70ex}{2ex}} \medskip}

\section*{Introduction}

Contextuality is one of the main features of Quantum Mechanics (QM), a feature which has been present since the early discussions of the founding fathers. However, there is still today no consensus in the community of physicists and philosophers who discuss the foundational problems of QM, about the exact definition or the physical meaning of quantum contextuality. In a recent paper \cite{Griffiths13}, Robert Griffiths has argued that ``Hilbert space QM is noncontextual''. According to Griffiths the problem of contextuality disappears if the apparatus is ``designed and operated by a competent experimentalist'' and we accept the {\it Single Framework Rule} ({\it SFR}). In the present paper we will argue that the conclussion derived by Griffiths is untenable. Firstly, we will argue that Kochen-Specker (KS) type contextuality has nothing to do with the measurement processes nor the measurement problem ---solving this problem does not necessarily remove contextuality. Secondly, that the consideration of counterfactual situations is endemic when reasoning by means of physics. Griffiths' {\it SFR} amounts to an unprincipled (or impractical) denial of this fact. Thirdly, that Griffiths' demonstration implies, as a direct consequence, that QM allows a {\it Global Valuation} for all properties independently of the context. This would mean, not only that KS theorem is false but also that the {\it SFR} is completely unnecessary. Fourthly, and finally, that quantum contextuality should be reconsidered ---instead as a problem that we need to escape--- as a main feature of the formalism which must be positively taken into account in order to produce a coherent representation of physical reality and explain what is QM talking about. 

The paper is organized as follows. In section 1, we provide a physical and philosophical analysis of the meaning of quantum contextuality. In section 2, we shall focus on the distinction between the basis problem and the measurement problem. Section 3 introduces the definition of {\it Meaningful Physical Statements (MPS)} and the need to consider {\it Counterfactual Reasoning (CR)} as a {\it necessary condition} in order to provide a coherent physical representation of reality. In section 4, we analyse and discuss Griffiths' arguments against quantum contextuality in terms of four main points presented in \cite{Griffiths13}. Finally, in section 5, we discuss what should be considered as a problem (and what not) in QM.

\section{Contextuality and Kochen-Specker Theorem}

Griffiths argues already in the abstract of his paper that:

\begin{quotation}
\noindent {\small``[...] quantum mechanics is noncontextual if quantum properties are represented by subspaces of the quantum Hilbert space (as proposed by von Neumann) rather than by hidden variables. In particular, a measurement using an appropriately constructed apparatus can be shown to reveal the value of an observable $A$ possessed by the measured system before the measurement took place, whatever other compatible ([$B,A$] = 0) observable $B$ may be measured at the same time.''  \cite[p. 174]{Griffiths13}}
\end{quotation}

\noindent This definition of contextuality discussed by Griffiths in \cite{Griffiths13} is a widespread generalization of the KS theorem which has become quite popular within the foundational literature. We have traced this exposition of KS theorem back to Peres' excellent book {\it Quantum Theory: Concepts and Methods} \cite[Chapter 7]{Peres02}, where he presents this explanation of contextuality as a direct consequence of the KS theorem. We leave a deeper analysis of this explanation for a future work. For the purpose of the present paper it is enough to make explicit the fact that this way of defining contextuality is a direct consequence of the KS theorem. Thus, if the former is proven to be false, the latter must be also false.    

According to Griffiths [{\it Op. cit.}, p. 174]: ``a substantial literature has accumulated which would throw doubt on this or suggest the opposite on the basis of various arguments related to the Kochen-Specker theorem.'' Griffiths builds up his argumentation against contextuality from the analysis and discussion of four main points or ideas:

\begin{enumerate}
{\it \item[i.] The idea that physical reality according to QM is not classical reality.}

{\it \item[ii.] The idea that in order to understand QM one needs to use the {\it SFR} and abandon {\it CR}.}

{\it \item[iii.] The idea that one can ``get rid'' of Schr\"odinger's cat through the {\it SFR}.}

{\it \item[iv.] The idea that contextuality can be discussed through the analysis of quantum measurements.}
\end{enumerate}

\noindent We will return to a detailed analysis of Griffiths' arguments in section 4. Before, we would like to discuss in some detail the meaning of quantum contextuality according to the Kochen-Specker theorem which is what Griffiths' demonstration proves to be false. 

The general characterization of the idea of ``classical reality'' can be condensed in the notion of {\it Actual State of Affairs (ASA)}.\footnote{See \cite{RFD14} for a detailed discussion and definition of this notion in the context of classical physics.} This particular (metaphysical) representation was developed by physics since Newton's mechanics and can be formultaed in terms of {\it systems constituted by a set of actual (definite valued) preexistent properties.} From a realist viewpoint, such {\it preexistent ASA} is independent of epistemic measurements and observations ---which have the only purpose of ``discovering'' or ``unveiling'' such existent state of affairs. However, in the case of quantum theory we have serious difficulties to interpret the orthodox quantum formalism in terms of ``classical reality''. In fact, we have no set of adequate physical concepts that would allow us to understand what is going on according to the theory. Indeed, an evidence of the deep crisis of physical representation within the theory of quanta is the fact that more than one century after its creation the physics community has reached no consensus about what the theory is talking about. 

There are multiple standpoints and strategies that have been adopted in order to discuss the meaning of QM. For example, Fuchs and Peres \cite[p. 70]{FuchsPeres00} have argued, in a paper entitled {\it Quantum Theory Needs No `Interpretation'}, that ``[...] quantum theory does not describe physical reality. What it does is provide an algorithm for computing probabilities for the macroscopic events (``detector clicks'') that are the consequences of experimental interventions. This strict definition of the scope of quantum theory is the only interpretation ever needed, whether by experimenters or theorists.'' This instrumentalist perspective is satisfied with having an ``algorithmic recipe'' that allows us to calculate measurement outcomes from the formalism. As a consequence, there is no need to supplement the theory with an interpretation that would conceptually explain its relation to physical reality. 

In contradistinction, representational realism stresses the idea that in order to produce meaningful physical representations mathematical formalisms must be necessarily complemented with networks of physical concepts. From this viewpoint, it is simply not enough to claim that ``according to QM the structure of the world is like Hilbert space'' or that ``the quantum formalism predicts the correct measurement outcomes''. That is not doing the job of provding a conceptual representation of what is going on according to the theory. To explain in physics is to provide meaningful conceptual representations of empirically adequate mathematical structures. Thus, the task of the physicist and the philosopher of physics is to produce theories that provide adequate physical representations which allow us to grasp and understand the world around us beyond the reference to measurement predictions and mathematical structures. 
 
We all agree that QM has a rigorous formalism and is empirically adquate. The challenge of representational realism with respect to QM is to find a conceptual scheme which matches the orthodox formalism and explains what the theory is talking about. In this respect, there seems to be two main lines of research which attempt to provide an answer to this challenge. The first one investigates the possibility that QM makes reference to the same physical representation provided by classical physics; i.e. that it talks about an {\it ASA}. This is the main idea presupposed, for example, by the hidden variables program which, as noticed by Bacciagaluppi  \cite[p. 74]{Bacciagaluppi96}, attempts to ``restore a classical way of thinking about {\it what there is}.'' The second possibility is to consider that QM might describe physical reality in a different ---maybe even incommensurable--- way to that of classical physics. This possibility seems to be endorsed by Griffiths \cite[p. 174]{Griffiths13} who argues that many of the problems with the interpretation of QM come from ``the view that the real world is classical, contrary to all we have learned from the development of quantum mechanics in the twentieth century.'' In relation to this debate regarding the conceptual meaning of QM, one of the main aspects that divides the foundational community is the issue of quantum contextuality. 

Contextuality is directly related to the impossibility to interpret the orthodox formalism of QM in terms of an {\it ASA} \cite{deRonde11}. This impossibility is something completely new in physics which has always dealt ---at least, until QM--- with commutative structures. In classical physics, every physical system may be described exclusively by means of its \emph{actual properties}, taking ``actuality'' as expressing the \emph{preexistent} mode of being of the properties themselves, independently of observation ---the ``pre'' referring to its existence previous to measurement.\footnote{Notice that even relativity theory is founded in the description of an {\it ASA} constituted by `preexistent space-time events'.} The evolution of the system may be described by the change of its actual properties. Mathematically, the state is represented by a point $(p; q)$ in the correspondent phase space $\Gamma$ and, given the initial conditions, the equation of motion tells us how this point moves in $\Gamma$. Physical magnitudes are represented by real functions over $\Gamma$. These functions can be all interpreted as possessing definite values at any time, independently of physical observation. Thus, as mentioned above, each magnitude can be interpreted as being actually preexistent to any possible measurement without conflicting with the mathematical formulation of the theory. In this scheme, speaking about potential or possible properties usually refers to functions of the points in $\Gamma$ to which the state of the system might arrive to in a future instant of time. These points, in turn are also completely determined by the equations of motion and the initial conditions. 

In QM, contrary to the classical scheme, physical magnitudes are represented by operators on ${\cal H}$ that, in general, do not commute. This mathematical fact has extremely problematic interpretational consequences for it is then difficult to affirm that these quantum magnitudes are \emph{simultaneously preexistent} (i.e., objective). In order to restrict the discourse to  sets of commuting magnitudes, different Complete Sets of Commuting Operators (CSCO) have to be (subjectively) chosen. And here is where the mixing of the objective and the subjective takes place. Indeed, the way to solve this uncomfortable situation within the orthodox approach is to include a subjective choice ---of one context, between the many incompatible ones--- that reintroduces superficially the classical structure at the price of erasing counterfactual reasoning. It is this {\it ad hoc} interpretational move, what needs to be physically justified in case we attempt to provide an objective realist representation of QM. Unfortunately, what many consider the best candidate to account for this interpretational maneuver, namely, the principle of decoherence, has failed to provide a convincing physical explanation of the quantum to classical limit.\footnote{As remarked by Bacciagaluppi \cite{Bacc12}, some physicists and philosophers still believe that ``decoherence would provide a solution to the measurement problem of quantum mechanics. As pointed out by many authors, however (e.g. Adler 2003; Zeh 1995, pp. 14-15), this claim is not tenable. [...] Unfortunately, naive claims of the kind that decoherence gives a complete answer to the measurement problem are still somewhat part of the `folklore' of decoherence, and deservedly attract the wrath of physicists (e.g. Pearle 1997) and philosophers (e.g. Bub 1997, Chap. 8) alike.''} Regardless of the many claims in the opposite direction, there are serious doubts, specially within the foundational literature, that decoherence is able to produce a convincing explanation ---beyond practical purposes--- of how a single basis is objectively chosen through the quantum measurement process (see \cite{DawinThebault15, Kastner14}). 

In QM the frames under which a vector is represented mathematically are considered in terms of orthonormal bases. We say that a set $\{x_1,\ldots,x_n\}\subseteq {\cal H}$ an $n$-dimensional Hilbert space is an \emph{orthonormal basis} if $\langle x_{i} | x_{j} \rangle = 0$ for all $1 \leq i , j \leq n$ and $\langle x_i|x_i\rangle=1$ for all $i=1,\ldots,n$. A physical quantity is represented by a self-adjoint operator on the Hilbert space ${\cal H}$. We say that $\mathcal{A}$ is a $\emph{context}$ if $\mathcal{A}$ is a commutative subalgebra generated by a set of self-adjoint bounded operators $\{A_1,\ldots,A_s\}$ of ${\cal H}$. Quantum contextuality, which was most explicitly recognized through the Kochen-Specker (KS) theorem \cite{KS}, asserts that a value ascribed to a physical quantity $A$ cannot be part of a global assignment of values but must, instead, depend on some specific context from which $A$ is to be considered. Let us see this with some more detail.

Physically, a global valuation allows us to define the preexistence of definite properties. Mathematically, a  \emph{valuation} over an algebra $\mathcal{A}$ of self-adjoint operators on a Hilbert space, is a real function satisfying,

\begin{enumerate}
\item[1.] \emph{Value-Rule (VR)}: For any $A\in\mathcal{A}$, the value $v(A)$ belongs to the spectrum of $A$, $v(A)\in\sigma(A)$.
\item[2.] \emph{Functional Composition Principle (FUNC)}: For any $A\in\mathcal{A}$ and any real-valued function $f$, $v(f(A))=f(v(A))$.
\end{enumerate}

\noindent We say that the valuation is a \emph{Global Valuation (GV)} if $\mathcal{A}$ is the set of all bounded, self-adjoint operators. In case $\mathcal{A}$ is a context, we say that the valuation is a \emph{Local Valuation (LV)}. We call the mathematical property which allows us to paste consistently together multiple contexts of {\it LVs} into a single {\it GV}, {\it Value Invariance (VI)}. First assume that a {\it GV} $v$ exists and consider a family of contexts $\{ A_i \}_I$. Define the {\it LV} $v_i:=v|_{A_i}$ over each $A_i$. Then it is easy to verify that the set $\{v_i\}_I$ satisfies the \emph{Compatibility Condition (CC)}, 

$$v_i|_{ A_{i} \cap A_j} =v_j|_{A_i\cap A_j},\quad \forall i,j\in I.$$

\noindent The {\it CC} is a necessary condition that must satisfy a family of {\it LVs} in order to determine a {\it GV}. We say that the algebra of self-adjoint operators is \emph{VI} if for every family of contexts $\{ A_i\}_I$ and {\it LVs} $v_i: A_i \rightarrow \mathbb{R}$ satisfying the \emph{CC}, there exists a {\it GV} $v$ such that $v|_{A_i}=v_i$.

If we have {\it VI}, and hence, a {\it GV} exists, this would allow us to give values to all magnitudes at the same time maintaining a {\it CC} in the sense that whenever two magnitudes share one or more projectors, the values assigned to those projectors are the same in every context. The KS theorem, in algebraic terms, rules out the existence of {\it GVs} when the dimension of the Hilbert space is greater than $2$. The following theorem is an adaptation of the KS theorem ---as stated in \cite[Theorem 3.2]{DF}--- to the case of contexts:

\begin{thm}[KS Theorem] If ${\cal H}$ is a Hilbert space of $\dim({\cal H}) > 2$, then a global valuation is not possible.
\end{thm}

After having recalled the KS theorem we are now ready to add some physical discussion. Some remarks are in order: 

\begin{enumerate}

\item[I.]  {\it KS type theorems preclude a physical representation of the formalism in terms of an {\it ASA}.}  If physical reality is conceived in terms of an {\it ASA}, then it is not possible to claim that the quantum state, $\Psi$, describes an actual situation irrespectively of the choice of a context. Not all observables can be considered to be simultaneously actual (real). 

\item[II.]  {\it KS type theorems have nothing to do with probabilities.} Going against a common phrase ---continuously repeated within the literature--- that says that ``QM is a probabilistic theory'' it should be clear that KS theorem does not talk about mean values of observables ---as it is the case for Bell inequalities. KS theorem discusses about the definite values of quantum properties \cite[chapter 5]{deRonde11}.

\item[III.]   {\it KS type theorems have nothing to do with measurements.} There is no need of actual measurements for the KS theorem to stand. The theorem is not talking about measurement outcomes, but about the {\it preexistence} of properties. About the constraints implied by the formalism to projection operators (interpreted in terms of properties that pertain to a quantum system). Quantum contextuality cannot be tackled through an analysis in terms of measurements simply because there is no reference at all to any measurement outcome. 

\item[IV.]  {\it KS type theorems have nothing to do with the evolution or dynamics of properties.} There is no evolution or dynamics considered for the KS theorem to stand. The theorem makes reference to the possible values of projection operators (interpreted as properties of a system) at one single instant of time. There is no question regarding the evolution of such properties, it is only their simultaneous consistent values considered from different contexts which is at stake. 

\item[V.] {\it KS type theorems in a nutshell.} Put in a nutshell, quantum contextuality deals with the formal conditions that any realist interpretation which respects orthodox Hilbert space QM must consider in order to consistently provide an objective physical representation of reality.

\end{enumerate}

\section{Disentangling Orthodox Problems in QM}

The orthodox line of research deals with a specific set of problems which analyse QM from a classical perspective. This means that all problems assume as a standpoint ``classical reality'' and only reflect about the formalism in ``negative terms''; that is, in terms of the failure of QM to account for the classical representation of reality and its concepts: separability, space, time, locality, individuality, identity, actuality, etc. The ``negative'' problems are thus: {\it non-}separability, {\it non-}locality, {\it non-}individuality, {\it non-}identity, etc.\footnote{I am grateful to Bob Coecke for this linguistic insight. Cagliari, July 2014.} These problems begin their analysis considering the notions of classical physics, assuming implicitly as a standpoint the strong metaphysical presupposition that QM should be able to represent physical reality in terms of such classical notions. But in between the many problems that can be found in the literature there are two unsolved problems which show the main difficulties of QM to provide an objective account of (classical) physical reality. The first problem relates directly to the issue of contextuality and is called the ``basis problem'':\\ 

\noindent {\it {\bf Basis Problem (BP):} Given the fact that $\Psi$ can be expressed by multiple incompatible bases (given by the choice of a Complete Set of Commuting Observables) and that due to the KS theorem the observables arising from such bases cannot be interpreted as simultaneously preexistent, the question is: how does Nature make a choice between the different bases? Which is the objective physical process that leads to a particular basis instead of a different one?}\\

\noindent Once again, the {\it BP} is a way of discussing quantum contextuality in ``negative terms''. The problem already sets the solution through the specificity of its questioning. The problem presupposes that {\it there exists a path} from the ``weird'' contextual quantum formalism to a classical non-contextual experimental set up in which classical discourse holds. If one could explain this path through an objective physical process, then the choice of the experimenter could be regarded as well as being part of such an objective process ---and not one that determines reality explicitly. Unfortunately, still today the problem remains with no solution within the limits of the orthodox formalism. There is no physical representation of the process without the addition of strange {\it ad hoc} rules, unjustified mathematical jumps and the like. These rules ``added by hand'', not only lack any physical justification but, more importantly, also limit the counterfactual discourse of the meaningful physical statements provided by the theory (we will come back to this point in section 3).

A very different problem ---sometimes also mixed and partly confused with the {\it BP}--- is the so called ``measurement problem'' which deals explicitly with the superposition principle and takes as a standpoint a specific basis or context ---``framework'' in the terminology employed by Griffiths.\\

\noindent {\it {\bf Measurement Problem (MP):} Given a specific basis (context or framework), QM describes mathematically a state in terms of a superposition (of states). Since the evolution described by QM allows us to predict that the quantum system will get entangled with the apparatus and thus its pointer positions will also become a superposition,\footnote{Given a quantum system represented by a superposition of more than one term, $\sum c_i | \alpha_i \rangle$, when in contact with an apparatus ready to measure, $|R_0 \rangle$, QM predicts that system and apparatus will become ``entangled'' in such a way that the final `system + apparatus' will be described by  $\sum c_i | \alpha_i \rangle  |R_i \rangle$. Thus, as a consequence of the quantum evolution, the pointers have also become ---like the original quantum system--- a superposition of pointers $\sum c_i |R_i \rangle$. This is why the {\it MP} can be stated as a problem only in the case the original quantum state is described by a superposition of more than one term.} the question is why do we observe a single outcome instead of a superposition of them?}\\

\noindent The {\it MP} is also a way of discussing the quantum formalism in ``negative terms''. In this case, the problem concentrates in the justification of measurement outcomes. It should be remarked that the {\it MP} presupposes that the basis (context or framework) ---directly related to a measurement set up--- has been already determined (or fixed). Thus it should be clear that there is no question regarding the contextual character of the theory within this specific problem. The {\it MP} has nothing to do with contextuality.  

Now that we have clearly separated these two different problems some remarks are in order. The mix of subjective and objective appears both in the {\it BP} and the {\it MP}. In the first case, orthodoxy has argued extensively that in order to recover a ``classical set up'' or ``classical discourse'' about properties (and learn which properties are definite valued and which are not), one needs to {\it choose} a specific context ---and this is where the {\it SFR} of Griffiths enters the scene.\footnote{As we have argued extensively in \cite{RFD14} this is not the case for QM, the choice of a context does not transform quantum properties into classical ones.} But if the ``choice'' is not physically justified in terms of an objective process, or the internal symmetries of the formalism, the definition of reality given by the (chosen) subset of (the now) actual properties is obviously subjective ---it depends explicitly on the choice of the experimenter, or equivalently, in the employment of the {\it SFR}. The context (or framework) is not determined {\it prior} to the choice of the experimenter, which means that the context cannot be considered as {\it preexistent}. In short, an (objective) independent reality cannot be dependent on a (subjective) choice. In the second case, the mix of subjective and objective pops up in the choice of the recording of the experiment ---as Wigner clearly exposed with his famous friend. The problem here is the shift from the physical representation provided when the measurement was not yet performed (and the system is described in terms of a quantum superposition), to when we claim that ``we have found a single measurement outcome'' which is not described by the theory. Since there is no physical representation of ``the collapse'', the subject (or his friend) seems to define it explicitly. Here the mixture of objective and subjective is due to an incomplete description of the state of affairs within the quantum measurement process (or ``collapse''). 

To summarize, quantum contextuality and the {\it BP} deal with the incompatibility of sets of contexts or bases which are in turn interpreted in direct relation to experimental set ups and definite properties. There is no measurement process involved here but the question of how multiple contexts or bases can be physically represented in accordance to the formalism of QM. In this case, the experimenter enters the scene explicitly in order to choose a specific context. In contradistinction, the {\it MP} deals only with the process of measurement within a fixed experimental set up (or context). The experimenter enters the scene because there is no objective physical account of when exactly the ``collapse'' takes place. What is important to notice for our proposes is that while the {\it BP} deals explicitly with the problem of quantum contextuality (i.e., the relation between bases or contexts), the {\it MP} deals with the problem of describing a situation by means of a quantum superposition ---always within an already fixed context. Both problems are independent, the solution of the former does not imply the solution of the latter and {\it vice versa}.

\section{Two Necessary Conditions for Physical Representation}

In order to discuss and analyse physical interpretations of a theory one should first agree with respect to what should be considered as {\it Meaningful Physical Statements (MPS)} within that theory. Furthermore, the theory should be capable of representing physically the {\it MPS} it talks about. From a representational realist perspective, this must be done not only in formal mathematical terms but also in conceptual terms through the addition of appropriate physical notions.

\begin{dfn}
{\sc Meaningful Physical Statements:} If given a specific situation a theory is capable of predicting in terms of definite physical statements the outcomes of possible measurements, then such physical statements are meaningful relative to the theory and must be constitutive parts of the particular representation of physical reality that the theory provides. Measurement outcomes must be considered only as confirming or disconfirming the empirical adequacy of the theory. 
\end{dfn}
 
\noindent  We must remark that {\it MPS} are not necessarily related to a closed and coherent physical representation of a theory. {\it MPS} can be, in principle, predictive instrumentalist or operational statements about specific physical situations. It is in fact the task of both physicists and philosophers of physics to complement {\it MPS} with adequate physical concepts, allowing us to construct a coherent representation of what is going on accoording to that theory. 

The {\it MPS} of Newtonian mechanics are different from those of Maxwell's electromagnetism or Einstein's relativity theory. For example, the statement `the particle will fall with acceleration $9.8 \frac{m}{s^2}$' from classical mechanics, is mathematically and conceptually different from the statement `the electromagetic field produced by a charge $q$ is $E_q$' which pertains to Maxwell's theory. Each theory must produce not only its own specific {\it MPS} but also a specific set of physical notions that allows us to properly represent them and explain what these statements are really talking about. The paradox introduced by QM is that, even though we possess a numerous set of {\it MPS} such as, for example, `the spin of the quantum particle is + with probability 0.5' or `the atom has a probability of decaying of 0.7', we do not possess an adequate set of concepts which allow us to garsp these statements in the same way as we do in Newton's mechanics or Maxwell's theory. Let us be clear about this point, in QM we do not understand conceptually what is the meaning of `quantum particle', `spin', `atom' or `probability'. When asked about such notions we can only provide a mathematical explanation or recall the empirical success of the theory. This type of formal or empirical explanation hides the fact that we do not know what QM is really talking about. 

For the representational realist, {\it MPS} necessarily pertain to a physical discourse which requires counterfactual reasoning in order to produce an objective account of physical reality. {\it Counterfactual Reasoning (CR)} is used and analysed by different disciplines. In the case of logicians and philosophers, {\it CR} is studied in terms of Kripke semantics, or possible worlds semantics. Even though this logical approach to counterfactuals has become popular in philosophy of QM (e.g. \cite{Griffiths02}), it has never been popular among physicists in general. In fact, physicists have always used counterfactuals in a rather (undefined) intuitive way in order to discuss physical experience realted to an objective description of reality. Let us provide thus a general definition of counterfactual reasoning which attempts to consider the actual {\it praxis} of physicists themselves.

\begin{dfn}
{\sc Counterfactual Reasoning:} If the theory is empirically adequate then the {\it MPS} it provides must be related to physical reality through a conceptual scheme. The possibility to make {\it MPS} realted to an objective physical representation requires necesarily a counterfactual discourse. {\it MPS} are not necessarily statements about future events, they can be also statements about past and present events. {\it CR} about {\it MPS} comprises all actual and non-actual physical experience. {\it CR} is the objective condition of physical discourse. 
\end{dfn}

{\it CR} is an indispensable element within a physical discourse which attempts to discuss an objective representation of physical reality. Many of the most important debates in the history of physics have taken place in the imagination of physicists making use of physical representation and counterfactual discourse. In the 18th century, Newton and Leibniz imagined different physical situations in order to draw conclusions about classical mechanics. At the beginning of the 20th century, Einstein's famous {\it Gedankenexperiments} in relativity theory made clear that the notion of {\it simultaneity} in Newtoninan mechanics had to be reconsidered, producing a revolution in our understanding of space and time. During the 1920s Solvay meetings, Bohr and Einstein discussed in depth many {\it Gedankenexperiments} related to QM. Some years later, Schr\"odinger imagined a strange situation in which a (quantum) cat was `dead' and `alive' at the same time. More than 50 years had to pass in order to empirically test the existence of quantum superpositions allowing technicians and experimentalists to explore amazing possibilities for quantum information processing. Also Einstein, Podolsky, Rosen and Bell had to wait till the 80s for Aspect and his group in order to be certain that the hidden variable project ---with which they wanted to replace QM by a ``complete theory''--- was not going to work out without giving up either {\it reality} or {\it locality}. These few examples show the crucial role played by conceptual representation and {\it CR} within the {\it praxis} of physicists. 

If we assume a representational realist stance, the conceptual representation must be capable of conceptually explaining the {\it MPS} produced by the theory, it must also produce a discourse which respects {\it CR}. Without {\it CR} in physical discourse one cannot imagine objective reality nor experience beyond the here and now. For a representational realist, the power of physics is {\it CR} itself. It allows us to state that ``if I performed this (or that) experiment'' then ---if it is a {\it MPS}--- the physical theory would tell me that ``the outcome will be $x$ (or $y$)'', and I do not need to actually perform the experiment! I know what the result will be independently of actually performing the experiment or not. That is the whole point of being a realist about physics, that I trust the theory to be talking about a physical representation of reality. 

Physicists are accustomed to play with the counterfactual statements produced by a theory. {\it CR} in physical discourse has nothing to do with time, evolution nor dynamics, it has to do with the possibility of representing objective physical reality and experience. A physical theory allows me to make counterfactual statements about the future, the present or the past, just in the same way physical invariance in classical mechanics connects the multiple frames of reference without anyone actually being in any of them. {\it CR} is the discursive invariance with respect to any physical phenomena. In classical mechanics (or relativity theory), we do not need to actually {\it be} in a specific frame of reference to {\it know} what will happen in that specific frame, or a different one. We can imagine and calculate what will happen in each frame, we can physically represent them to ourselves and translate what will happen in each of them through the Galilean (or Lorentz) transformations. For a realist, that is the magic of physics. Once we believe to have an empirically adequate theory, we realists are committed to the representation provided by that theory. It is this trust which allows physicists to imagine situations which escape not only the spatial region in which they live ---imagining for example what might happen with a distant star or even a back hole---, but also the technical possibilities of their time ---as the previous examples clearly exemplify.

From the previous analysis, two {\it necessary conditions} for the physical representation of theories can be stated:

\begin{enumerate}
\item[{\bf Necessary Condition I.}] {\it Every physical theory must be capable of producing MPS which can be empirically tested. }

\item[{\bf Necessary Condition II.}] {\it Every physical theory must be capable of producing a coherent counterfactual physical discourse by complementing the MPS it produces with adequate mathematical and conceptual schemes.}
\end{enumerate}

\section{Revisiting Griffiths' Arguments Against Quantum Contextuality}

In this section we attempt to show that Griffiths' arguments, which support the main ideas discussed in section 1 contain several flaws, mainly due to multiple shifts from the ontic to the epistemic level of analysis. The criticisms provided in this section can be easily extended to several (supposedly) realist interpretations of QM.

\subsection{``Quantum Reality'' Is Not ``Classical Reality''}

Griffiths argues that the ``wrong assumption'', that QM talks about ``classical reality'', has led to the idea that QM is contextual:

\begin{quotation}
\noindent {\small``How have so many come to [the] conclusion [that QM is contextual]? By adopting, we shall argue, the view that the real world is classical, contrary to all we have learned from the development of quantum mechanics in the twentieth century. In particular, discussions couched in terms of hidden variables typically assume that they are classical rather than the sort of thing one might expect in a quantum mechanical world.'' [{\it Op. cit.}, p. 174]}
\end{quotation}

\noindent As we have discussed above, KS theorem can be read as making exactly this same point against classical physical representations of QM: because of contextuality QM cannot be described in terms of a classical {\it ASA} kind of representation ---which is what amounts to ``classical reality''. In this sense, KS should be understood as an {\it ad absurdum} proof of the failure of the notion of (classical) {\it actual preexistence} to account for what QM is talking about. The conclusion that must be drawn is the following: in case we want to stay close to orthodox Hilbert space QM, then projection operators cannot be interpreted in terms of actually preexistent properties ---which is what, as we shall see, Griffiths ends up trying to do. 

According to Griffiths: 

\begin{quotation}
\noindent {\small``The way our approach avoids any conflict with the Kochen-Specker theorem is by denying {\it Re} [Realism about properties]. The claim that every observable possesses a value at every time is, indeed, inconsistent with a representation of quantum properties by subspaces of a Hilbert space. Consider, for example, a spin-half particle. There are distinct rays in the two-dimensional Hilbert space corresponding to $S_x = + \frac{1}{2}$ (in units of $h$) and to $S_x = - \frac{1}{2}$; also rays corresponding to $S_z = + \frac{1}{2}$ and $S_z = - \frac{1}{2}$, but there is no ray that can represent a simultaneous value of $S_x$ and $S_z$. {\it Students are told, correctly, that $S_x$ and $S_z$ cannot be measured simultaneously, and they ought to be told that the reason for this is that there is nothing there to be measured.} The projectors corresponding to $S_z$ do not commute with the projectors corresponding to $S_x$, and once one has accepted the connection between quantum properties and Hilbert subspaces proposed by von Neumann it makes no sense to speak of a spin half system in which, for example, $S_x = + \frac{1}{2}$ at the same time that $S_z = + \frac{1}{2}$.'' [{\it Op. cit.}, p. 179] (emphasis added)}
\end{quotation} 

\noindent As we have discussed above, the condition {\it Re} is perfectly well founded in the physical representation provided by classical Newtonian mechanics in terms of an {\it ASA} (see section 1). Now, the fact that two observables do not commute does not only imply that the observables cannot be measured together (epistemic incompatibility of measurements) but more importantly, that one cannot assign a {\it Global Valuation} to them (ontic incompatibility of preexistent properties) ---which precludes the possibility of representing what is going on in terms of an {\it ASA}. 
In this respect, it is important to distinguish the {\it ontic} incompatibility of properties from the {\it epistemic} incompatibility of measurements.

\begin{dfn}
{\sc Epistemic Incompatibility of Measurements:} Two contexts are epistemically incompatible if their measurements cannot be performed simultaneously. 
\end{dfn}

\begin{dfn}
{\sc Ontic Incompatibility of Properties:} Two contexts are ontically incompatible if their formal elements cannot be considered as simultaneoulsy preexistent. 
\end{dfn}

\noindent The fact that even in classical physics we can find epistemically incompatible measurement situations has been analysed by Diederik Aerts in \cite{Aerts83}. Aerts discusses the example of a piece of wood which might posses the properties of being ``burnable'' and ``floatable''. Obviously, both properties are {\it testable} through mutually incompatible experimental arrangements. Indeed, in order to measure the ``burnability'' we need to burn the piece of wood, but then ---becuase it has been burned--- it is not possible anymore to see if it floats. In order to measure the ``floatability'' we need to put the piece of wood in the water, but then ---becuase it is now wet--- its burnability cannot be tested. These two experiments are {\it epistemically incompatible}. 

However, the properties are not {\it ontically incompatible}, the epistemic realm of measurements does not make any direct reference to the ontic level of properties. In fact, in classical physics, all properties can be thought to exist as actual (ontic) properties. The formal Boolean structure of propositions allows an interpretation of them in terms of an {\it ASA}. This example shows that epistemic incompatibility of measurements does not imply the ontic incompatibility of preexistent properties. In fact, we know that all observables in classical mechanics can be regarded as ontically compatible due to the existence of {\it Global Valuations} of all observables (or properties) pertaining to any situation. Going back to Griffiths claim, the fact that measurements cannot be performed simultaneusly is an epistemic feature which does not involve a direct ontic consequence. One cannot jump ---if one is a realist--- from an epistemic statement such as ``[quantum observables] cannot be {\it measured} simmultaneously'', to an ontic conclusion such as: ``then there {\it is} nothing there''.

Griffiths continues arguing that:  

\begin{quotation}
\noindent {\small``It is somewhat odd that this particular principle [{\it Re}] should be identified with realism, since at the present time all available experimental evidence is in accord with Hilbert space quantum mechanics, and not with classical physics when the two disagree. If a Hilbert space provides the appropriate mathematics to describe everything from the quarks to the quasars, where in the real world, the one we live in, is there any part that satisfies the condition of `realism' given by {\it Re}?  It would be much less confusing if whenever `realism' were used in this way the adjective `classical' were prepended. The hidden variables of typical hidden variable theories are classical hidden variables, and it is for this reason that the attempt to use them for interpreting quantum theory has given rise to numerous conflicts with the latter.'' [{\it Op. cit.}, p. 179]}
\end{quotation}

\noindent We completely agree with Griffiths that the realism presupposed within the debate of quantum contextuality is in fact ``classical realism''. However, we disagree that one can, from a realist stance, argue only in terms of epistemic evidence. That should be clearly not enough for a realist ---like Griffiths himself claims to be. What is lacking is not a formalism that accounts for measurement outcomes. We already have that. What is lacking is a network of physical concepts which provides a representation that coherently relates to such formalism and explains {\it what the outcomes are talking about}. Just in the same way that in classical mechanics we understand that `a point in phase space' (in the formal level) relates to `a particle in Newtonian space-time' (in the conceptual level), or just like Maxwell's equations (in the formal level) relate to electromagnetic waves in space-time (in the conceptual level); we ---representational realists--- need to explain in detail which are the physical notions (in the conceptual level) that relate to the orthodox Hilbert space quantum structure (in the formal level). Clearly, classical entities (e.g. `particles' and `waves') are not doing the job. It is neither clear to the author of this paper that Griffiths' consistent histories interpretation is providing such non-classical representation of QM. 

As we mentioned above, if one is a representational realist who admits that having a formalism is not enough in order to provide a physical representation of a theory, there seems to be two different paths to follow. Either we develop a new (non-contextual) formalism that describes things in terms of an {\it ASA} (i.e., in terms of ``classical reality'') as the hidden variables program attempts to do; or we must find a new (non-classical) conceptual network which takes into account all {\it MPS} in QM ---even those of incompatible observables. Griffiths seems to claim that the problem of describing quantum reality has been already solved when he argues that: ``Hilbert space provides the appropriate mathematics to describe everything from the quarks to the quasars, where in the real world, the one we live in''. On the contrary, for the representational realist, a rigorous mathematical scheme and the empirical adequacy of a theory are {\it necessary} but {\it not sufficient} conditions in order to generate a physical representation. Physical representation also requires a network of adequate physical notions. Until we do not find such appropriate physical concepts we will not be able to say what QM is really talking about.

\subsection{Applying the {\it SFR} (Instead of {\it CR}) in QM}

In several papers, Griffiths ---who is certainly not alone in supporting this idea--- has argued against {\it CR} in physical discourse \cite{Griffiths02, Griffiths12, Griffiths13, Griffiths13b}. According to him \cite[p. 178]{Griffiths13}: ``Analyzing counterfactual questions is a bit tricky even in situations not entangled with quantum mysteries.'' Indeed, this idea of {\it CR} being something ``tricky'' or ``weird'' is quite widespread within the literature. In order to escape the consequences of {\it CR} Griffiths proposes to ``block'' this discursive condition in QM through the introduction of what he calls the {\it Single Framework Rule} ({\it SFR}).\footnote{Equivalent {\it ad hoc} rules like the {\it SFR} are provided in many interpretations of QM. For example, in the perspectival version of the modal interpretation proposed by Bene and Dieks \cite{BeneDieks02}, in Bub's bohmian version of the modal interpretation \cite{Bub97} or the contextual approaches by Karakostas \cite{Karakostas14} and Svozil \cite{Svozil14}.}

\begin{quotation}
\noindent {\small``The single framework rule asserts that any sort of discussion of the quantum system must be carried out in some framework of the sort just discussed, which is typically chosen because it has some events which are interesting for some reason or another. The physicist is free to choose any framework he pleases for describing the world; what the single framework rule prohibits is combining frameworks. [...] The well-known Kochen-Specker paradox is constructed precisely by forming a bridge, or one might better say bridges, between incompatible frameworks in such a way that one eventually ends up with a contradiction. {\it The histories approach disposes of the paradox by declaring the bridging invalid.}'' [{\it Op. cit.}, p. 180] (emphasis added)}
\end{quotation}

To ``declare the bridging invalid'' amounts to declare that the {\it CR} commonly used by physicists is invalid in quantum theory. {\it CR} is the discursive invariance which allows one to claim that in case we performed an experiment the phenomenon $A$ will have the result $a$ while the phenomenon $B$ will have the result $b$, and so forth. {\it CR} presupposes ---according to the realist stance--- that there is an objective (independent of subjective choices) physical representation of reality according to the theory. The fact that we can talk about such physical representation without the need to actually perform an experiment is the very condition of possibility for objective (conterfactual) discourse within physics. The bridging which Griffiths ``declares invalid'' introduces a choosing subject and gives rise to the basis problem discussed above in detail (section 2). However, it is important to remark that the probabilistic statements of QM do not require the choice of a context.\footnote{Given a vector in Hilbert space, $\Psi$, the Born rule allows us to predict the average value of (any) observable $O$, $\langle \Psi| O | \Psi \rangle = \langle O  \rangle$. This prediction is context (or basis) independent.}  The choice of the context is required only in order to seemingly recover a classical (Boolean) discourse about definite valued properties; or in other words, about ``classical reality''. 

Continuing his argumentation, Griffiths makes the point that: 

\begin{quotation}
\noindent {\small``It is important to note that the single-framework rule does not state that there is only one framework which can be used for a valid quantum description of a situation. {\it The quantum physicist is free to choose any framework}, consistent with the Hilbert space structure of quantum mechanics (and, in the case of histories, satisfying consistency conditions if the extension of the Born rule is to be used to assign probabilities), in order to describe a quantum system. {\it In some frameworks a particular observable A may possess some value, while in other frameworks it may not.} The existence of the latter does not preclude the possibility or the validity of a framework which include the former. The single framework rule is not a restriction on the use of frameworks; it is instead a prohibition against combining incompatible frameworks.'' [{\it Op. cit.}, p. 180] (emphasis added)}\end{quotation}

Some remarks are in order. Firstly, objective physical representation involves the need to be able to discuss of all contexts simultaneously. Properties of different contexts cannot be brought into reality by virtue of a choice if we want to claim at the same time that our representation of reality is not defined subjectively. There should be no subjective choice in order to claim that a given property is objectively actually existent or not (i.e., has a definite value or not). Secondly, as we argued above, the {\it ontological incompatibility} of definite values of quantum properties should not be confused with the {\it epistemological incompatibility} of distinct measurements. Due to the specific questioning I might not be able to perform two measurements simultaneously. As already remarked, in general, not all properties of a classical physical object can be measured at the same time. However ---and this is what really matters--- in classical physics all properties in a given situation are consistent with possessing a definite value; making it possible to understand any classical situation in terms of an objective {\it ASA}. Due to the Boolean structure of classical mechanics, all classical properties of a system are ontologically compatible. It is not the epistemological incompatibility of measurements ---to which Girffiths makes reference--- which is at stake in KS contextuality but the ontological incompatibility of the values of properties which allows a {\it Global Valuation} ---and consequently a description in terms of an objective {\it ASA}.  

Griffiths' argumentation seems to be in tension with the argument presented in the previous subsection. Griffiths now argues that the {\it SFR} should be used in order to talk about ``properties with preexistent values''. But this is exactly the ``classical'' physical representation that Griffiths had firmly criticized. The {\it SFR} seems to be an {\it ad hoc} rule only justified by the (supposedly) necessary import of the classical discourse. Let us summarize. On the one hand, introducing classical discourse has the only purpose to restore the classical representation of physics in terms of definite valued properties. But this goes against Griffiths earlier claim that QM cannot be described in terms of ``classical reality''. On the other hand, if QM would be non-contextual, as Griffiths claims to have demonstrated in \cite{Griffiths13}, the {\it SFR} would become completely unnecessary. These two lines of argumentation seem flagrantly in contradiction. Why should someone who claims that ``QM is not talking about classical reality'' impose {\it ad hoc} rules to QM which have the only purpose of restoring a (classical) discourse about ``classical reality''? Indeed, Griffiths seems to have created a paradoxical situation for his own approach. If he is correct in claiming that ``a measurement [...]can be shown to reveal the value of an observable $A$ possessed by the measured system before the measurement took place''\footnote{[{\it Op. cit.}, p. 174].} independently of the context, then there would exist a definite value for all observables (since $A$ can be any observable). But this is equivalent to saying that there exists a {\it Global Valuation} for all properties independently of the context. This would prove KS to be false which is an untenable result; since this theorem is a direct consequence of the quantum formalism mathematically proved in many different ways (see \cite{Peres02}). Furthermore, it would also turn Griffiths' {\it SFR}  completely unnecessary. On the contrary, if KS is correct, then Hilbert space QM is contextual and the value of $A$ will differ when measured together with $B$ or $C$ (with [$B,A$] = 0, [$C,A$] = 0, but $[B,C] \neq 0$) \cite[p. 196]{Peres02}.

\subsection{``Getting Rid'' of Quantum Superpositions}

In his paper, Griffiths [{\it Op. cit.}, p. 177] considers the question: ``What will occur if the experimenter prepares an initial state $|\psi \rangle = \frac{1}{\sqrt{2}}(|a_1\rangle + |a_2\rangle)$ which is a superposition corresponding to two distinct eigenvalues of $a_1$ and $a_2$ of $A$ and then carries out a measurement?'' Griffiths argues that: 

\begin{quotation}
\noindent {\small``[...] a properly constructed quantum measurement apparatus does just what it was designed to do, measure properties of a system corresponding to a particular decomposition of the identity. It will do this for a system initially prepared in one of the states it was designed to measure, and for one initially prepared in a superposition of those states. To discuss the latter one needs to employ probabilities, in particular the (extended) Born rule, both in order to have a measurement pointer with a well-defined position, and to show that this position is appropriately correlated with one of the properties the apparatus was designed to measure.'' [{\it Op. cit.}, p. 178]}
\end{quotation} 

In section 2 we made clear that solving the {\it MP} has no relation at all to the solution of the {\it BP} or contextuality. The solution of the former does not imply in any way the solution of the latter. A superposition is a particular mathematical representation of a vector $\Psi$ in Hilbert space which implies the choice of a particular basis. The {\it SFR} cannot help in dissolving or doing away with quantum superpositions because quantum superpositions are defined within a fixed context (or framework). Let us remark this important point:  there is no quantum superposition composed by states which give rise to {\it incompatible observables}. It is a mathematical fact which goes beyond interpretation that superpositions are mathematical expressions within a fixed basis; i.e. a complete set of {\it commuting observables}. The criticism provided by Griffiths to the many worlds interpretation of QM also makes clear the fact there is a mix between the {\it MP} and the {\it BP} in his argumentation against Schr\"odinger's cat. 

\begin{quotation}
\noindent {\small``[...] the histories approach is distinctly different from the Everett or many-worlds interpretation with its insistence that `the wave function', in this instance the state onto which $V$ projects, represents fundamental physical reality. From the histories point of view the difficulties which many-worlds advocates have in explaining how ordinary macroscopic physics can be consistent with their perspective is not unrelated to the fact that they are seeking to assign simultaneous reality to properties which in the quantum Hilbert space are represented by incompatible projectors.'' [{\it Op. cit.}, p. 177]}
\end{quotation}

\noindent Contrary to this claim, many worlds focuses in solving the {\it MP}, and deals in no way with incompatible projectors. The many worlds interpretation attempts to provide physical meaning to quantum superpositions by multiplying classical reality as many times as terms are found in a specific superposition (described in relation to a particular context). The original solution provided by many worlds to the {\it MP} does not deal in any way with contextuality nor the {\it BP}. 

Finally, we should remark that a representational realist should not feel uncomfortable with the non-classical features expressed by quantum superpositions. Instead of considering quantum superpositions in terms of a ``mathematical tool'' that calculates pre-probabilities, she should be trying to develop a coherent physical notion which is capable of making sense of this most important formal element of the theory. If we suspect that QM cannot be represented in terms of ``classical reality'',  then it neither makes sense to interpret quantum superpositions in terms of (classical) definite valued properties. Something we know is untenable due to the fact quantum probability cannot be interpreted in terms of ignorance.

\subsection{Measurement Process and Contextuality}

The discussion provided by Griffiths continues calling the attention to the need of considering the measurement process in the discussion about contextuality. Griffiths then argues that: 

\begin{quotation}
\noindent {\small``[...] getting rid of the ghost of Schr\"odinger's cat, we still need to show that the measurement apparatus actually carries out a measurement; i.e., the outcome pointer position is properly correlated with a previous property of the measured system. For this purpose we need an extension of Born's rule that allows probabilities to be assigned to a closed quantum system at three or more times, and this in turn requires the use of consistent (or decoherent) families of histories.'' [{\it Op. cit.}, p. 177]}
\end{quotation}

\noindent Once again, the discussion is shifted to the justification of the measurement process. 
According to Griffiths: ``The most direct approach to determining whether quantum theory is or is not contextual is to analyse the process that goes on in a quantum measurement.'' [{\it Op. cit.}, p. 176] We have argued extensively why this cannot be the case. Quantum measurements have nothing to say about the contextual character of QM simply because contextuality says nothing about measurements. From a realist perspective, (epistemic) measurements cannot be considered as more fundamental than the (ontic) physical representation provided by the theory in terms of its mathematical structure and the physical network of concepts. For a representational realist about physics, measurement outcomes are only part of a verification procedure. As Griffiths himself remarks in his book \cite[p. 361]{Griffiths02}:  ``If a theory makes a certain amount of sense and gives predictions which agree reasonably well with experimental or observational results, scientists are inclined to believe that its logical and mathematical structure reflects the structure of the real world in some way, even if philosophers will remain permanently skeptical.''

\section{The Contextuality Problem}

The main idea behind the representational realist stance is that when a mathematical formalism is coherently related to a network of physical concepts, it is possible to produce a physical representation of reality which allows us to describe particular physical phenomena. It is not enough to say that QM cannot be explained in terms of ``classical reality''. Neither is enough to argue that, ``according to QM the world has the struture of Hilbert space''. That is not doing the job of explaining in conceptual terms what QM is talking about. QM has proven already to be empirically adequate and the orthodox formalism is mathematically rigorous. So instead of changing the formalism or adding {\it ad hoc} rules in order to restore a classical discourse and representation, there is a different strategy that could be considered. 

According to this new strategy what needs to be done is to construct a new net of (non-classical) physical concepts capable of interpreting the formalism {\it as it is}. In this respect, contextuality should not be regarded as a ``ghost'' which we need to fight or destroy, it should be accepted as one of the key features that must help us in understanding the type of physical reality implied by QM. Technological and experimental developments are going much faster than the present discussions regarding foundational issues about QM. These discussions seem to be still stuck by the limits imposed by ``classical reality'' and the ``no-problems'' discussed in section 2. This is one of the main reasons we believe that the field is in need of a strong criticism. The {\it BP} and {\it MP} which attempt to build a bridge between QM and our classical understanding of reality, could be replaced in the literature by new problems which accept the possibility of developing a new representation of physical reality according to QM. In \cite{deRonde11}, we put forward two new problems which could help us to think, from a different perspective, the problem of interpretation in QM. The problem that interests us here should replace the {\it BP}.\\ 

\noindent {\it {\bf Contextuality Problem:} Given the fact that Hilbert space QM is a contextual theory, the question is: which are the concepts that would allow us to coherently interpret the formalism and provide a representation of physical reality that accounts for this main feature of the theory in a natural way?}\\
 
\noindent Contextuality is a consequence of the mathematical structure of QM, a formal scheme which has been allowing us for more than one century to produce the most outstanding physical predictions. The feature of contextuality emerges from the orthodox formalism of QM itself, it is not something external to it. Unfortunately, instead of regarding quantum contextuality as a new interesting feature of a new physical theory, many attempts in the literature have centred their efforts in attempting to restore a representation of QM in terms of an {\it ASA}. The consistent histories interpretation, just like in the case of the hidden variables approach, also denies the formalism in order to restore a classical discourse. They do so, not by changing it explicitly ---as in the case of the former--- but by adding unnecessary rules (e.g. the {\it SFR}) to account for the meaningful physical statements the theory already provides ---without such rules. We believe that to deny contextuality just because it obstructs an interpretation of the theory in terms of actual (definite valued) properties would be tantamount to trying to deny the Lorentz transformations in special relativity simply because of its implications to the contraction of rigid rods. Indeed, this was the attempt of most conservative physicists until Einstein made the strong interpretational move of taking seriously the formalism of the theory and its phenomena, and derived a new net of physical notions in order to coherently understand the new theory. 

If we accept the orthodox formalism, then contextuality is the crux of QM. It is contextuality that which needs to be physically interpreted, instead of something that needs to be destroyed because of its non-classical consequences. The contextuality problem opens the possibility to truly discuss a physical representation of reality which goes beyond the classical representation of physics in terms of an {\it ASA}. We are convinced that without a replacement of the problems addressed in the literature there is no true possibility of discussing an interpretation of QM which provides an objective non-classical physical representation of reality. We know of no reasons to believe that this is not doable.

\section*{Acknowledgments} 

I wish to thank R. B. Griffiths for private discussions regarding the subject of quantum contextuality and physical reality. I also want to thank G. Domenech, N. Sznajderhaus and M. Graffigna for a careful reading of previous versions of this manuscript. Finally, I would like to thank two anonymous referees (3 and 4) who through their comments and remarks allowed me to improve the manuscript substantially. This work was partially supported by the following grants: FWO project G.0405.08 and FWO-research community W0.030.06. CONICET RES. 4541-12 (2014-2015).


\end{document}